\documentclass[a4paper]{jpconf}
\usepackage{graphicx}
\usepackage[usenames]{color}
    \def\|{\partial}
    
    \def\Oo {\displaystyle}

\begin{document}

\noindent
doi: 10.1088/1742-6596/973/1/012007

\noindent
http://iopscience.iop.org/article/10.1088/1742-6596/973/1/012007/pdf

\title{Dynamics of unstable sound waves in a non-equilibrium medium at the nonlinear stage \footnote{Khrapov S., Khoperskov A. Dynamics of unstable sound waves in a non-equilibrium medium at the nonlinear stage // Journal of Physics: Conference Series, 2018, v.973, 012007} }

\author{S.S. Khrapov$^*$, A.V. Khoperskov$^{**}$}

\address{Volgograd State University, Volgograd, 400062, Russia}

\ead{$^*$khrapov@volsu.ru, $^{**}$khoperskov@volsu.ru}

\begin{abstract} %

A new dispersion equation is obtained for a non-equilibrium medium with an exponential relaxation model of a vibrationally excited gas.
We have researched the dependencies of the pump source and the heat removal on the medium thermodynamic parameters.
The boundaries of sound waves stability regions in a non-equilibrium gas have been determined.
The nonlinear stage of sound waves instability development in a vibrationally excited gas has been investigated within CSPH-TVD and MUSCL numerical schemes using parallel technologies OpenMP-CUDA.
We have obtained a good agreement of numerical simulation results with the linear perturbations dynamics at the initial stage of the sound waves growth caused by instability.
At the nonlinear stage, the sound waves amplitude reaches the maximum value that leads to the formation of shock waves system.

\end{abstract}

\section{Introduction}
The dynamics of sound waves in thermodynamically nonequilibrium media  significantly differs from the equilibrium media acoustics \cite{1Bauer1973,2Kogan1985,3Kogan1986,4Molevich2008,5Zavershinskiy2014}. Examples of such media are the vibrationally excited gas, the nonisothermal plasma, the chemically active mixtures, etc.
In the non-equilibrium media the second (bulk) viscosity and the acoustic dispersion may turn negative.
 The negative acoustic dispersion denotes that the equilibrium (low-frequency) sound velocity $c_0$ becomes greater than the ``frozen'' (high-frequency) sound velocity $ c_\infty $.
The medium is acoustically active when the negative second viscosity gives the main contribution to the total viscosity value.
Various aspects of this problem have been discussed in Ref.\cite{Galimov2006}.
In such gas the increase in the sound waves amplitude due to the instability may lead to a shock waves system formation~\cite{6Makaryan2005}.

Conditions for sound instability caused by a special viscosity may arise in accretion disks around relativistic compact objects (white dwarfs, neutron stars and black holes) or in circumstellar gas-dust disks.
The aim of such astrophysical studies is an attempt to construct an accretion model of rotating gas using turbulence phenomenological models~\cite{a1-Khrapov1999,a2-Nedugova2003,a3-Yao2007}.

In Ref. \cite{6Makaryan2005} the mechanisms of nonequilibrium gas medium influence on the structure of small amplitude shock waves caused by new dispersion-viscosity properties were considered.
It was also shown there that a finite area perturbation in a nonequilibrium medium decays with the formation of autowave pulses and periodic autowave.

It should be noted that research of essentially nonlinear stage of sound waves and strong shock waves dynamics in nonequilibrium media requires the numerical gasdynamic methods of high accuracy.
In current paper, we have numerically simulated the nonlinear stage of unstable sound waves development in non-equilibrium media, in which gas molecules are excited by vibrational degrees of freedom stipulated by the energy of the external pump source.

As a rule, the numerical schemes for the hydrodynamics equations are being developed and tested for gas in the thermodynamic equilibrium state.
Therefore, studying the nonlinear structures (shock waves and rarefaction waves) in a nonequilibrium gas we used the numerical algorithms based on various approaches of continuous medium description (Lagrangian and Eulerian).
Since the Euler and Lagrangian numerical schemes have different dispersion properties, our combined approach allows to identify unphysical numerical solutions that may appear in non-equilibrium medium models.

\section{Materials and Methods}

\subsection{Mathematical model}
A nonequilibrium (acoustically active) medium may be simulated using the vibrationally excited gas equations along with the exponential relaxation model \cite{6Makaryan2005}:
$$
\frac{\partial \varrho}{\partial t} +  {\rm div}(\varrho \mathbf{v})=0\,, \eqno(1)
$$
$$
\frac{\partial \varrho \mathbf{v}}{\partial t} +  {\rm div}(\varrho \mathbf{v} \mathbf{v})=-\nabla p +  \varrho \,\mathbf{f}\,, \eqno(2)
$$
$$
\frac{\partial E}{\partial t} +  {\rm div}[(E+p)\mathbf{v}]=\varrho\left(Q - I +\mathbf{v}\mathbf{f}\right)\,, \eqno(3)
$$
$$
\frac{\partial E_\nu}{\partial t} +  {\rm div}(E_\nu\mathbf{v})=\left(\frac{E_{\nu e} - E_\nu}{\tau} + \varrho \, Q \right)\,, \eqno(4)
$$
where $\varrho$ is the gas density, $p$ is the gas pressure, $\mathbf{v}$ is  the velocity vector, $\mathbf{f}$ is the specific external force, $Q$, $I$ is the pump and heat sink power, respectively, $E_\nu$ is the energy of the vibrational degrees of freedom of the excited gas, $E_{\nu e}$ is the equilibrium value of the vibrational energy, $\tau$ is the vibrational relaxation time.
The expression of total energy for a vibrationally excited gas can be represented as follows
$$
 E=\frac{p}{\gamma -1} + \frac{\varrho |\mathbf{v}|^2}{2} + E_\nu\,, \eqno(5)
$$
where $\gamma$ is the adiabatic exponent.

In general the quantities of $Q$, $I$ and $\tau$ depend on the thermodynamic parameters of the medium (the density $\varrho$ and temperature $T$). The functions $Q(\varrho, T)$, $I(\varrho, T)$ and $\tau(\varrho, T)$ in the vicinity of the local equilibrium values of $\tau_0, Q_0, I_0, \varrho_0, T_0$ we represent as a power series expansion:
$$
\tau = \tau_0 \left(\frac{\varrho}{\varrho_0}\right)^{\tau_\varrho} \left(\frac{T}{T_0}\right)^{\tau_T}\,, \quad
Q = Q_0 \left(\frac{\varrho}{\varrho_0}\right)^{Q_\varrho} \left(\frac{T}{T_0}\right)^{Q_T}\,, \quad
I = I_0 \left(\frac{\varrho}{\varrho_0}\right)^{I_\varrho} \left(\frac{T}{T_0}\right)^{I_T}\,. \eqno(6)
$$
Dependences (6) allow to investigate the dynamics of sound waves in a variety of acoustically active media by selection of appropriate parameters $\tau_\varrho, \tau_T, Q_\varrho, Q_T, I_\varrho, I_T$.

The equation of state is applied to close the system of equations (1)--(6)
$$
p = R \varrho \,T\,, \eqno(7)
$$
where $R$ is the gas constant.

\subsection{Linear stability analysis}
Let us conduct the standard procedure of linearizing the system of equations (1)--(4) in the one-dimensional case and represent all the quantities in the form $f=f_0+\widetilde{f}$.
Using the WKB approximation and substituting a solution in the form of plane harmonic waves $\widetilde{f}(t,x)=\widehat{f}(\omega, k)\exp\{-i\omega t + i k x\}$ (where $\omega$ is the natural frequency, $k$ is the wave number) into the obtained linearized equations, we obtain a system of linear algebraic equations (SLAE).
Equating the determinant of SLAE with zero and taking into account Eqs. (5)--(7), we get the dispersion equation for an acoustically active medium:
$$
\omega^2\left[(A-i\delta)+i\frac{(1-i\delta)\gamma_1}{\delta R}(I_0 I_T - Q_0 Q_T)\right] =
$$
$$
k^2 c_0^2 \left[(B-i\delta)+i\frac{(1-i\delta)\gamma_1}{\omega\gamma T_0 R}(I_0(I_T-I_\varrho)-Q_0(Q_T-Q_\varrho))\right]\,, \eqno(8)
$$
where $\delta=\omega\tau_0$, $\gamma_1 = \gamma - 1$, $c_0=\sqrt{\gamma R T_0}$ is the equilibrium low-frequency sound velocity in the gas, $A=1+(C_k+S(\tau_T+Q_T))\gamma_1$, $\displaystyle B=1+(C_k+S(\tau_T-\tau_\varrho+Q_T-Q_\varrho))\gamma_1/\gamma$,
$\displaystyle C_k = \frac{\partial E_{\nu e}}{R\partial T}$, $\displaystyle S=\frac{\tau_0 Q_0}{R T_0}$ is a degree of non-equilibrium of the medium.

The dispersion relation (8) is a generalization of the equation obtained in Ref.\cite{6Makaryan2005} to the case $Q_\varrho \neq 0, Q_T \neq 0, I_\varrho \neq 0, I_T \neq 0$.
For certain values of the parameters $\tau_\varrho, \tau_T, Q_\varrho, Q_T, I_\varrho, I_T$ and degree of medium non-equilibrium,  $S$, corresponding to the negative second (bulk) viscosity and acoustic dispersion, solutions with acoustic increment (${\rm Im} k < 0$) appear in Eq.~(8).

For acoustically active medium described by Eq. (8), four characteristic nonequilibrium regions can be distinguished depending on the degree of nonequilibrium $S$ \cite{6Makaryan2005}:
\begin{enumerate}
  \item An area with positive acoustic dispersion and second viscosity ($0<S<S_1$). In this region sound waves are damped like in the equilibrium medium with viscosity.
  \item An area with negative acoustic dispersion and second viscosity ($S_1\leq S < S_2$). In this region unstable sound waves with ${\rm Im} k < 0$ exist. For $S \rightarrow S_2$ the equilibrium sound velocity is substantially higher than the high-frequency (``frozen'') velocity.
  \item An area with positive dispersion and negative bulk viscosity ($S_2\leq S < S_3$). In this area a low-frequency sound cannot propagate due to $c_0^2/c_\infty^2 < 0$.
  \item An area with positive dispersion and negative bulk viscosity ($S_3\leq S$). In this region, along with acoustic instability, a thermal instability is possible.
\end{enumerate}

In accordance with (8), the values of $S_1, S_2, S_3$ have the form:
$$
S_1 = - \frac{\gamma_1 C_k}{\tau_\varrho + \gamma_1 \tau_T}\,, \quad
S_2 = - \frac{\gamma_1 C_k}{(\tau_\varrho+I_\varrho) + \gamma_1 (\tau_T + I_T)}\,,
$$
$$
S_3 = - \frac{I_T - Q_T + (I_\varrho - Q_\varrho)(C_k + 1/\gamma_1)}
{\tau_T(I_\varrho - Q_\varrho)-\tau_\varrho(I_T - Q_T) + Q_T I_\varrho - Q_\varrho I_T}\,.
$$

We adopt following values of the parameters: $\gamma=1.4$, $C_k=0.1$, $S=0.1$, $\delta_\varrho = -1$, $\delta_T = -2$, $Q_\varrho=Q_T=I_\varrho=I_T=0$ as the basic parameters corresponding to a typical laser CO$_2$--containing medium under normal conditions.

The analysis of solutions of the dispersion equation (8) indicates an increase in the acoustic increment with decreasing of $\tau_\varrho, \tau_T, I_\varrho, I_T$ and increasing of $Q_\varrho, Q_T$.


\subsection{Numerical nonlinear model}

To construct the numerical model, we use the standard procedure for discretization of a continuous medium on a space-time grid defined by the grid points $(x_j = x_{j-1}+h, y_k=y_{k-1}+h, t_n = t_{n-1}+\Delta t)$ and for each function we have $f(x,y,t) \rightarrow f(x_j, y_k, t_n,)=f_{j,k}^n$.

We use two numerical methods: CSPH-TVD (Combined Smoothed Particle Hydrodynamics -- Total Variation Diminishing, \cite{Khrapov2011,khrapov2013mech}) and MUSCL (Monotone Upwind Scheme for Conservation Laws, \cite{9-Khoperskov2016,10-Khoperskov2013}), which are based on different approaches to the continuous medium description.
Let's consider each of them in more detail below.

The CSPH-TVD method contains two main stages.
At the first Lagrangian stage, a modified SPH algorithm is used \cite{Monaghan2005,Khrapov-Khoperskov-2017!SuperDays}.
The explicit method of Godunov type \cite{7-Toro2012} and the TVD principle underlie the second Eulerian stage.
For the non-equilibrium gas on the basis of equations (1)--(4), the general calculation scheme is:
$$
\mathbf{U}_{j,k}^{n+1}  =  \widetilde{\mathbf{U}}_{j,k}^{n+1}  -
\frac{\Delta t}{h}\left(\mathbf{F}_{j+1/2,k}^{n+1/2} - \mathbf{F}_{i-1/2,k}^{n+1/2} + \mathbf{G}_{j,k+1/2}^{n+1/2} - \mathbf{G}_{j,k-1/2}^{n+1/2}\right) \,, \eqno(9)
$$
where
$$
\mathbf{U} = \left(
               \begin{array}{c}
                 \varrho \\
                 \varrho u \\
                 \varrho v \\
                 E \\
                 E_\nu \\
               \end{array}
             \right)\,, \qquad
\mathbf{F} = \left(
               \begin{array}{c}
                \varrho u\\
                \varrho u^2\\
                \varrho u v\\
                u E\\
                u E_\nu
\end{array}
             \right)\,, \quad
\mathbf{G} = \left(
               \begin{array}{c}
                \varrho v\\
                \varrho v u\\
                \varrho v^2\\
                v E\\
                v E_\nu
\end{array}
             \right)\,.
$$

In equation (9), the quantity $\widetilde{\mathbf{U}}_{j,k}^{n+1}$ is calculated on the Lagrangian stage using a two-step time recalculation (predictor-corrector), providing the second order of accuracy in time:

\textit{Predictor}
$$
 \widetilde{\mathbf{U}}_{j,k}^* = \mathbf{U}_{j,k}^n +
        \Delta t\, \mathbf{Y}_{j,k}\left(\mathbf{U}^{n},\, \mathbf{r}^{n}\,\right)  \, ,  \qquad
        \mathbf{r}_{j,k}^*  =  \mathbf{r}_{j,k}^{n}  +  \Delta t\frac{\mathbf{v}_{j,k}^n + \mathbf{v}_{j,k}^*}{2}\, , \eqno(10)
$$

\textit{Corrector}
$$
 \widetilde{\mathbf{U}}_{j,k}^{n+1} =\frac{\mathbf{U}_{j,k}^n + \widetilde{\mathbf{U}}_{j,k}^*}{2}  +
        \frac{\Delta t}{2}\, \mathbf{Y}_{j,k}\left(\mathbf{U}^{*},\, \mathbf{r}^{*}\,\right)  \, ,  \qquad
        \mathbf{r}_{j,k}^{n+1}  =  \frac{\mathbf{r}_{j,k}^{n} + \mathbf{r}_{j,k}^{*}}{2} +
        \frac{\Delta t}{2}\frac{\mathbf{v}_{j,k}^n + \mathbf{v}_{j,k}^{n+1}}{2}\, , \eqno(11)
$$
where $\mathbf{r}$ is the radius vector of the position of the ``liquid'' SPH-particles, which simulates a continuous medium at the Lagrangian stage.
At each time point $t_n$, the particles are in the center of the Euler cells before the start of the predictor-corrector procedure.
The quantity $\mathbf{Y}_{j, k}$ entering into equations (10)--(11) is equal to
$$
    \mathbf{Y}_{j,k} = -\left(
    \begin{array}{c}
        0 \\
        \Oo  \varphi_{j,k}\sum\limits_{j'=j-1}^{j+1}\sum\limits_{k'=k-1}^{k+1}\,\varphi_{j',k'} \, \nabla W_{j,k} +
             \varrho_{j,k} \mathbf{f}_{j,k}    \\
        \Oo \varphi_{j,k}\sum\limits_{j'=j-1}^{j+1}\sum\limits_{k'=k-1}^{k+1}\,\varphi_{j',k'}
         \frac{\mathbf{v}_{j,k}+\mathbf{v}_{j',k'}}{2} \nabla W_{j,k} +
         \varrho_{j,k}\left(Q_{j,k} - I_{j,k} + \mathbf{v}_{j,k}\mathbf{f}_{j,k}\right)    \\
        \Oo \varrho_{j,k}\left(\frac{E_{\nu e} - E_{\nu\, j,k}}{\tau_{j,k}} + Q_{j,k}\right)    \\
    \end{array}
    \right)\,, \eqno(12)
$$
where $\varphi = \sqrt{2 p}$, $W_{j,k}=W(|\mathbf{r}_{j,k} - \mathbf{r}_{j',k'}|,h)$ is the smoothing kernel used to approximate the spatial derivatives in equations (1)--(4) in the SPH approach.

The values of the fluxes $\mathbf{F}_{j \pm 1/2, k}^{n+1/2}$ and $\mathbf{G}_{j, k \pm 1/2}^{n+1/2}$ are computed on the cell boundaries using solutions of the Riemann problem at time $t_{n+1/2}$.
Approximate methods of Lax-Friedrichs (LF), Harten-Lax-van Lier (HLL), and the modified Harten-Lax-van Lier method (HLLC) \cite{8-Toro1999} are most often used to solve the Riemann problem.
In this paper, we use the HLL method for the flux $\mathbf{F}_{j+1/2, k}^{n+1/2}$
$$
\mathbf{F}_{j + 1/2, k}^{n+1/2} = \left\{
  \begin{array}{ll}
    \mathbf{F}_L, & 0 < S_L \\
    \Oo \frac{S_R \mathbf{F}_L - S_L \mathbf{F}_R + S_L S_R \left(\mathbf{U}_R - \mathbf{U}_L\right)}{S_R - S_L},
      & S_L \leq 0 \leq S_R \\
    \mathbf{F}_R, & 0 > S_R
  \end{array}
\right.\,, \eqno(13)
$$
where $S_L = \min (u_L - c_L, u_R - c_R)$, $S_R = \max (u_L + c_L, u_R + c_R)$, and the indices $L, R$ determine the flow parameters to the left and right of the boundary $j+1/2$, respectively.

The construction of the second-order accuracy scheme together with the time predictor-corrector algorithm requires the use of piecewise-linear approximation of $\mathbf{U}$ values inside cells.
In the CSPH-TVD method, the values of the flux parameters on the left ($L$) and on the right ($R$) from the boundary $j+1/2$ have the following form:
$$
    \begin{array}{c}
        \Oo \mathbf{U}^L  =   \mathbf{U}_{j,k}^{n+1/2} +
        \frac{1}{2}\,\left(h - \xi^{\,n+1}_{j,k}\right)\, \mathbf{\Theta}_{j,k}^{n}\, ,
        \\    \\
        \Oo \mathbf{U}^{R}  \,=\,   \mathbf{U}_{j+1,k}^{n+1/2}  - \frac{1}{2}\,
        \left(h + \xi^{\,n+1}_{j+1,k}\right)\, \mathbf{\Theta}_{j+1,k}^{n}\, ,
    \end{array}
 \eqno(14)
$$
where $\xi^{\,n+1}_{j,k}$ is the displacement of the ``liquid'' particle relative to the cell center ($j, k$) at time $t_{n + 1}$.
Since the slopes $\mathbf{\Theta}$ of the piecewise linear dependence (14) must satisfy the TVD-condition, it is necessary to apply the limiter functions.

The MUSCL scheme \cite{9-Khoperskov2016,10-Khoperskov2013} also uses the two-step predictor-corrector method, the piecewise linear approximation with TVD-limiters and the HLL method for solving the Riemann problem.
For the MUSCL method, equation (9) takes the form:
$$
\mathbf{U}_{j,k}^{n+1}  =  \mathbf{U}_{j,k}^{n}  -
\frac{\Delta t}{h}\left(\mathbf{F}_{j+1/2,k}^{n+1/2} - \mathbf{F}_{i-1/2,k}^{n+1/2} + \mathbf{G}_{j,k+1/2}^{n+1/2} - \mathbf{G}_{j,k-1/2}^{n+1/2}\right) + \Delta t \,\mathbf{Y}_{j,k}^{n+1/2}\,, \eqno(15)
$$
Since the expressions for the fluxes and source terms in equation (15) differ from the corresponding expressions for CSPH-TVD, we write them explicitly as
$$
\mathbf{F} = \left(
               \begin{array}{c}
                \varrho u\\
                \varrho u^2+p\\
                \varrho u v\\
                u (E+p)\\
                u E_\nu
\end{array}
             \right)\,, \quad
\mathbf{G} = \left(
               \begin{array}{c}
                \varrho v\\
                \varrho v u\\
                \varrho v^2+p\\
                v (E+p)\\
                v E_\nu
\end{array}
             \right)\,, \quad
\mathrm{Y} = \left(
               \begin{array}{c}
                0\\
                \varrho f_x\\
                \varrho f_x\\
                \varrho (\mathbf{v} \mathbf{f} + Q - I)\\
                \varrho((E_{\nu e} - E_\nu)/\tau + Q)
\end{array}
             \right)\,.
$$

For the MUSCL method, we set $\xi = 0$ in equation (14) for constructing piecewise linear dependencies.

New combined methods based both on the Euler (grid method) approach and on the Lagrangian (particle motion) approach, began to develop in recent years.
The combined TVD + SPH method demonstrated high efficiency for problems of the numerical simulation of long waves runup on a shore \cite{Shokin2016}.
The hybrid code SPH-TVD was used in astrophysical gas dynamics to simulate the stellar disruption by a massive black hole \cite{Lee1996}.

In general, the stability condition for the explicit numerical schemes CSPH-TVD and MUSCL~is:
$$
\Delta t < \min\limits_{j,k} \left(  \frac{h}{2\,|\mathbf{v}_{j,k}|} \, , \frac{h}{|\mathbf{v}_{j,k}|+c_{j,k}} \right)
\, , \eqno(16)
$$

To increase the speed of gas-dynamic simulation on high-resolution grids, we parallelized the numerical algorithms CSPH-TVD and MUSCL using OpenMP/CUDA technologies and performed calculations on a Computing Clusters with NVIDIA Tesla K40 and K80 coprocessors.

We normalize all physical quantities using $\varrho_0$, $p_0$, $\tau_0$ and $\sqrt{p_0/ \varrho_0}$ as units of density, pressure, time and velocity, respectively.
And then we will use dimensionless quantities only.

\subsection{Testing numerical model}

\begin{figure}[!ht]
    \centering
    \includegraphics[width=0.9\textwidth]{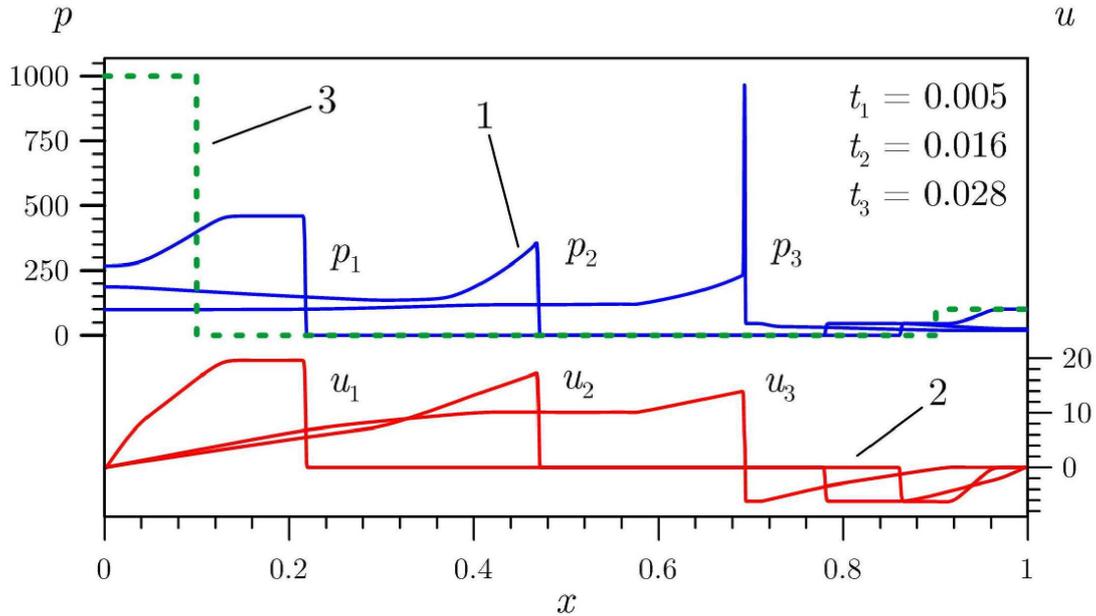}
    \vskip 0mm
    \caption{The interaction of two shock waves is shown at different times.
    The solid blue lines on the upper panel are the pressure distributions $p(x)$ and the red lines are the velocities
    $u(x)$ on the bottom panel at different times.
    The initial pressure profile $p(x,t = 0)$ is represented by the green dotted line.}
    \label{fig:Blast-wave1}
\end{figure}

\begin{figure}[!ht]
    \centering
    \includegraphics[width=0.5\textwidth]{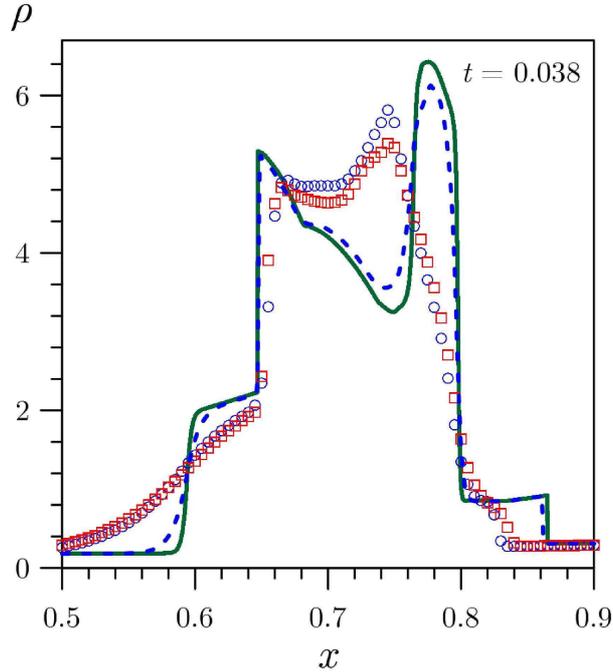}
    \vskip 0mm
    \caption{The result of a collision of two shock waves.
    The blue circles are the density after the collision ($t = 0.038$) for the CSPH-TVD scheme with $N = 200$ and the dotted blue line is the model with $N = 1600$; the red squares are the result for MUSCL ($N = 200$).
    The exact solution is shown by the solid line.}
    \label{fig:Blast-wave2}
\end{figure}

Let's consider the test of the collision of two shock waves~\cite{Woodward-Colella}.
The initial state consists of three regions filled with an ideal gas with an adiabatic index $\gamma=1.4$ and the computational domain is bounded by solid walls.
The initial density is homogeneous ($\rho(x)=1$, $u(x)=0$, $x \in [0,1]$), but we set two pressure discontinuities:
$$
p(x)   =   \left\{
             \begin{array}{ll}
               1000, & \hbox{$0 \le x < 0.1$;} \\
               0.01, & \hbox{$0.1 \le x \le 0.9$;} \\
               100, & \hbox{$0.9 < x \le 1$.}
             \end{array}
           \right.
$$
As a result, two strong convergent shock waves, two contact density discontinuities and two diverging discharge waves are formed.
Figure~\ref{fig:Blast-wave1} shows the dynamics of interaction of two shock waves for $N = 10^3$.
At $t_1 = 0.028$ the shock waves collide with the formation of a characteristic peak at the collision position $x = 0.7$ and as a result two divergent shock waves are formed (Figure~\ref{fig:Blast-wave2}).

Figure~\ref{fig:Blast-wave2} shows the results obtained by the CSPH-TVD and MUSCL schemes at the time $t = 0.038$ after the collision of shock waves.
Both methods demonstrate a good resolution of the contact discontinuity formed as a result of the collision of convergent shock waves, and the width of the fronts for the generated divergent shock waves are approximately equal.
The numerical solution on a very detailed grid for $N = 10^5$ is shown as an exact solution.
Convergence of our numerical solutions is clearly noticeable with increasing spatial resolution in discretization.

\section{Results and Discussion}
For numeric simulation of the acoustic waves dynamics in a nonequilibrium medium, we specify forced oscillations from a source with initial conditions $u = 0$, $\varrho = 1$, $p = 1$. To calculate the vibrational relaxation time, we define the values of parameters $\tau_\varrho = -1$, $\tau_T = -1.5$, $Q_\varrho = Q_T = I_\varrho = I_T = 0$.
The wavelength of the generated perturbations is $\lambda = 0.1$. We set the value of the degree of medium nonequilibrium to be $S=0.5$.
\begin{figure}[!ht]
\vskip  0.0\hsize \hskip 0.0\hsize
  \centering
  \includegraphics[width=0.8\hsize]{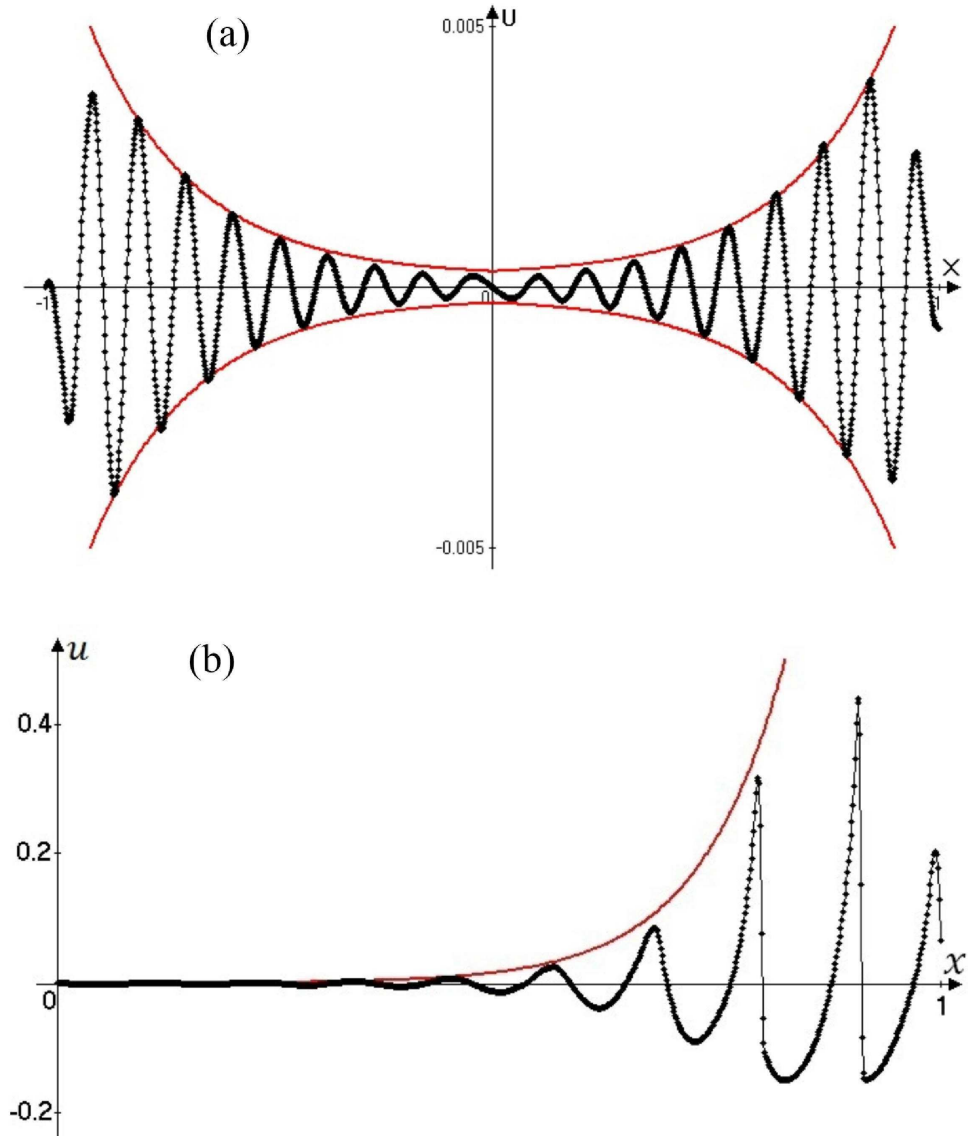}
\vskip  0.0 mm
  \caption { The distribution of the velocity perturbation: a) at the linear stage at time $t = 0.8$ and $N = 800$;
  b) at the nonlinear stage at time $t = 0.8$ and $N = 1600$ (the section of the calculation region $0\leq x \leq 1$ is shown.}
\label{fig:1Dsound-wave}
\end{figure} %

Consider the initial linear stage of sound waves propagation. Within the linear analysis, the perturbation amplitude grows according to the law $\Oo \sim \exp{\frac{-{\rm Im} k }{\lambda }x}$.
For the environmental parameters given above Figure~\ref{fig:1Dsound-wave}a shows the results of numerical gasdynamic simulations with acoustic increment ${\rm Im} k = - 0.43$. As Figure~\ref{fig:1Dsound-wave}a demonstrates in the linear stage the exponential growth of the perturbation amplitude corresponding to a linear stability analysis (solid red line) is well tracked.

We set $S=1$, ${\rm Im} k=-1.034$ to study the dynamics of sound waves at the nonlinear stage. For given medium parameters Figure~\ref{fig:1Dsound-wave}b shows the distributions of velocity perturbation in the sound wave. At time $t=0.8$, the initial harmonic perturbations turn into a system of strong shock waves. The amplitude, velocity and shape of such disturbances are determined only by the nonequilibrium medium parameters.

\begin{figure}[!ht]
\vskip  0.0\hsize \hskip 0.0\hsize
  \centering
  \includegraphics[width=0.3\hsize]{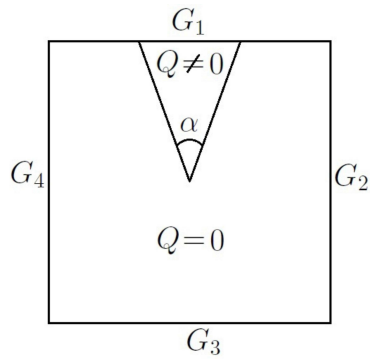}
\vskip  0.0 mm
  \caption {The computational domain scheme (the active medium is given by the sector with the angle
  $\alpha = 30^\circ$).}
\label{fig:2Dschem}
\end{figure}

\begin{figure}[!ht]
\vskip  0.0\hsize \hskip 0.0\hsize
  \centering
  \includegraphics[width=0.999\hsize]{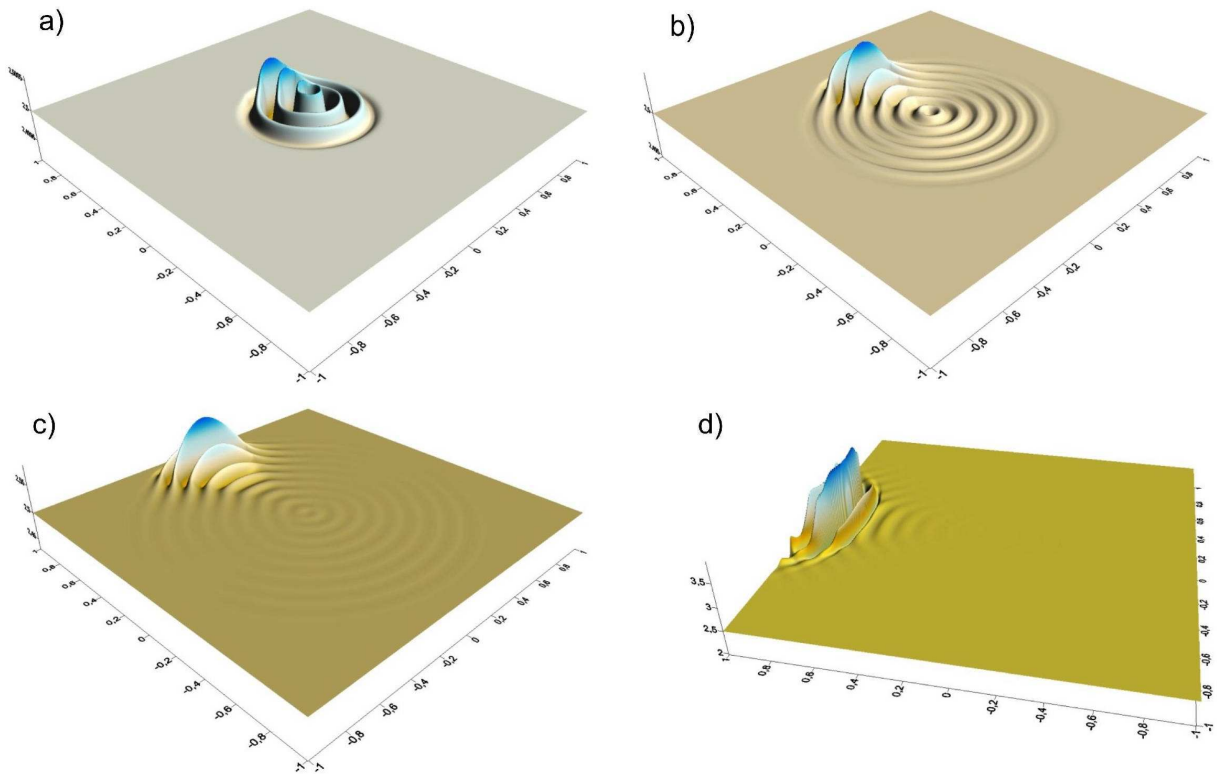}
\vskip  0.0 mm
  \caption { The density perturbation distribution at different moments of time: a) $t = 0.25$, b) $t = 0.5$, c) $t = 0.75$, d) $t = 1$.}
\label{fig:2Dsound-wave}
\end{figure}

Consider the problem of simulation of the dynamics of acoustic waves in two-dimensional space.
The active medium is set in some part of the computational domain, at a triangle with an angle $\alpha = 30^\circ$
(see Figure~\ref{fig:2Dschem}).
The non-percolation condition (a solid wall) is set at the boundary $G_1$, and the free flow conditions are specified at the boundaries $G_2, G_3, G_4$. According to such problem formulation the source begins to generate perturbations diverging in all directions, but the acoustic waves amplitude increases only in the sector where the non-equilibrium medium is determined (see Figure~\ref{fig:2Dsound-wave}).

At time $t=1$, the wave reaches the boundary $G_1$ and is reflected (see Figure~\ref{fig:2Dsound-wave}d).
The calculation continues that demonstrates a good stability and balance of the numerical scheme for reflection problems of a strong shock waves system.

Thus, the results obtained above indicate a significant effect of stationary non-equilibrium on the sound waves evolution.
 In the active media sound waves are unstable, so a strong increase of the acoustic disturbance amplitude is observed. In this case, the wave packet passes through all the stages of the instability development: linear and nonlinear stages in the formation of strong shock waves system.

\section{Conclusion} 
In the article we have researched the sound waves steadyness in a nonequilibrium acoustically active medium.
Using a linear steadyness analysis, we have obtained a new dispersion equation accounting for the following functional dependences: $\Oo \tau \propto \varrho^{\tau_\varrho}T^{\tau_T}$, $\Oo Q \propto \varrho^{Q_\varrho} T^{Q_T}$  $\Oo I \propto \varrho^{I_\varrho} T^{I_T}$ for the vibrational relaxation time, pump power and heat sink, respectively.
The influence of parameters $\tau_\varrho$, $\tau_T$, $Q_\varrho$, $Q_T$, $I_\varrho$, $I_T$ on the acoustic increment has been studied in detail. The boundaries of the steadyness regions in a nonequilibrium vibrationally excited gas have been determined as well.

Numerical algorithms using the CSPH-TVD and MUSCL methods have been developed to study the nonlinear stage of unstable acoustic waves development in a nonequilibrium vibrationally excited gas. The parallel computing technologies OpenMP-CUDA have been implemented to accelerate the calculations.

At the initial stage of sound waves propagation in a nonequilibrium medium, the results of numerical gas dynamic modeling are in good agreement with the linear steadyness analysis, which indicates the adequacy of the developed algorithms.
At the nonlinear stage of development unstable sound waves form a shock waves system which maximum amplitude depends on the degree of the medium nonequilibrium $S$ and model parameters $\tau_\varrho$, $\tau_T$, $Q_\varrho$, $Q_T$, $I_\varrho$, $I_T$.

Applying the numerical gas dynamic methods of simulation, the dynamics of cylindrical sound waves which propagates simultaneously in acoustically active and inactive media has been investigated.
Due to the presence of interface between two media, the shock waves formed at the nonlinear stage of the acoustic instability development have an inhomogeneous structure along the azimuthal coordinate.

\section*{Acknowledgments}
The second author has been supported by the Ministry of Education and Science of the Russian Federation (government task No.\,2.852.2017/4.6).
The authors are thankful to the RFBR (grants 16-07-01037, 15-02-06204).
The research is carried out using the equipment of the shared research facilities of HPC computing resources at Lomonosov Moscow State University.

\section*{References}

\providecommand{\newblock}{}

\end{document}